\documentclass[a4paper]{jpconf}
\usepackage{graphicx}
\begin{document}
\title{Neutrino Interaction Measurements Using the T2K Near Detectors}

\author{Daniel G. Brook-Roberge for the T2K Collaboration}

\address{University of British Columbia, Vancouver, BC, Canada}

\ead{droberge@triumf.ca}

\begin{abstract}
  The T2K near detectors provide a rich facility for measuring neutrino
  interactions in a high-flux environment. This talk will discuss the
  near detector CC-inclusive normalization analysis for the T2K
  oscillation result in detail, along with the present result, and
  describe the plan for its extension to more sophisticated
  measurements. Selection criteria for CCQE interactions will be
  presented, as will a strategy for calculating cross-section difference
  between plastic scintillator and water. The unique capacities of the
  near detectors to measure other exclusive CC and NC channels in a
  narrow-band off-axis beam will also be explored.
\end{abstract}

\section{Introduction}

The T2K (Tokai to Kamioka) long-baseline neutrino oscillation
experiment\cite{t2knim} aims to make high-precision measurements of
oscillation parameters using a beam of muon neutrinos. In particular,
the appearance of electron neutrinos in the beam is studied in an
effort to measure a non-zero value of the mixing parameter
$\theta_{13}$. Electron neutrino appearance provides a particular
challenge for the near detector, as the electron neutrino fraction in
the muon neutrino beam is the most significant background to this
measurement. 

The T2K experiment has had two experimental runs; one at beam power up
to 50 kW from January to June 2010, and one at beam power up to 150 kW
from November 2010 to March 2011, when it was cut short by the Tohoku
earthquake. $1.43\times10^{20}$ protons on target were accumulated by T2K prior
to the earthquake.

Recently, T2K has published an electron appearance result with
suggestion of large $\theta_{13}$ ($\sin^22\theta_{13} \approx
0.1$)\cite{t2kres}. The significance of this result has driven further
interest in improving background estimates using the near detector.

\section{The T2K Experiment}

The T2K experiment has three main components: the beam source, the
near detector complex, and the far detector, Super-Kamiokande. The
ND280 near detector is the focus of this paper.

\subsection{Beamline}

The T2K beam is produced from the 30 GeV proton beam of the J-PARC
Main Ring. These protons are extracted in eight bunches every 3
seconds and collided with a long graphite target. The positively
charged secondary particles are focussed down a 96 m decay volume
using three magnetic horns. The predominant contribution is from
pions, which decay $\pi^+ \rightarrow \mu^+\nu_\mu$ to produce muon
neutrinos. All charged decay products are stopped in a beam dump while
the neutrinos pass through freely. The electron neutrino component is
produced by other particles decaying in the beam pipe, such as muons
and kaons.

The T2K beamline is angled 2.5$^\circ$ away from Super-Kamiokande;
this off-axis beam technique provides a narrower energy spectrum
peaking near the oscillation maximum, and dramatically reduces the
high-energy tail. T2K is the first oscillation experiment to use the
off-axis configuration.

\subsection{Far Detector (Super-Kamiokande)}

The 50 kt water \v{C}erenkov detector Super-Kamiokande (Super-K) is
the far detector for T2K, sitting 295 km from the production target at
J-PARC. Super-K images the \v{C}erenkov rings produced by charged
particles from neutrino charge-current interactions to detect incident
neutrinos. The ring imaging allows a high-purity separation of
electrons and muons, as the two particles produce \v{C}erenkov rings
with much differently defined edges. 

Many of the Super-K backgrounds are intended to be well constrained by
ND280. In addition to the intrinsic electron neutrino fraction of the
T2K beam, we have substantial background from neutral pions. A neutral
pion decays to two photons, which are indistinguishable in Super-K
from electrons. If one of the decay photons escapes or is otherwise
missed, the remaining photon will mimic a CCQE electron neutrino
interaction, i.e.~our appearance signal.

\subsection{Near Detectors}

Two near detectors are located in a pit 280 m from the production
target at J-PARC. The INGRID detector is a cross-shaped arrangement of
14 iron/scintillator neutrino detector modules, centred on the beam
axis. The event rates in these modules are used to measure the stability
of the beam flux and direction. At the T2K design intensity, the
interaction rate should be sufficient to monitor these on a day-to-day
basis.

\begin{figure}
  \centering
  \includegraphics[width=2.25in]{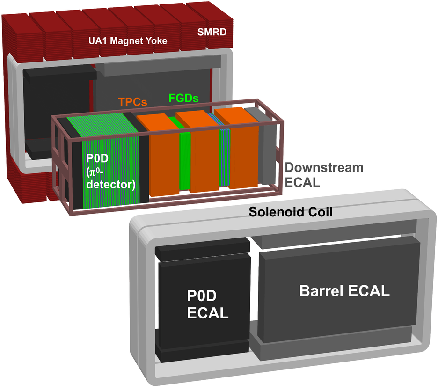}
  \caption{An exploded view of the ND280 detector.}
  \label{fig:nd280}
\end{figure}

At the off-axis angle to the neutrino beam is the ND280
spectrometer. The ND280 detector consists of two central detectors,
the P0D (Pi-Zero Detector) and the Tracker, surrounded by an
electromagnetic calorimeter (ECal) and placed within the 0.2 T
magnetic field of the UA1 magnet. Interleaved in the magnet yokes is
an additional detector, the Side Muon Range Detector (SMRD). A full
diagram of ND280 can be found in Figure \ref{fig:nd280}.

The Tracker consists of three large-volume Time Projection Chambers
(TPCs) and two Fine Grained Detectors (FGDs). The TPC gas detectors
allow high precision 3D tracking of charged particles and measurements
of the energy loss, while the plastic scintillator FGDs provide an
active target mass ($\approx 1.1$ tons per FGD) allowing tracking back
to the neutrino interaction vertex. The TPC can measure the momentum
and charge of particles from their curvature in the magnetic field,
and provides a measure of $dE/dx$ for particle identification. Tracks
from the TPC can then be traced back into the FGD and associated with
FGD activity, allowing the vertex to be localized and the fiducial
volume established. FGD-only tracking can then be used to find
short-range particles such as protons and pions. 

The P0D aims to measure neutral current $\pi^0$ production and takes
the form of a hybrid calorimeter/tracking detector. The central region
of the P0D alternates scintillator planes, water targets, and brass
sheets for observation of tracks and showers, while the ends use a
scintillator/lead combination to contain the full showers within the
P0D. 

The lead-scintillator ECal aims to identify particles escaping from
the central detector and to detect photons produced in the inner
detector. For the first T2K run in early 2010, only the ECal module at
the downstream end of the Tracker was operational. The remainder of
the ECal was installed and ready for the second run starting November 2010.

\section{Current Results}

Two neutrino interaction results are available from ND280, both from a
Tracker-based analysis: a measurement of the $\nu_\mu$ charge-current
interaction rate and a measurement of the $\nu_e$ fraction. These
analyses attempt to use the Tracker to find neutrino interactions in
the FGD. Presently only inclusive charge-current measurements are
complete; work is ongoing to measure specific channels.

\subsection{Tracker $\nu_\mu$ Rate Measurement}

The current $\nu_\mu$-based analysis is a comparison of the Tracker
muon neutrino interaction rate with a NEUT Monte Carlo prediction. The
data/Monte Carlo ratio is used to scale the flux prediction at Super-K
and provides a useful contribution to the oscillation measurement. We
select CC $\nu_\mu$ events by selecting tracks that begin in the FGD
fiducial volume and make it into a downstream TPC. Incoming events are
removed by requiring no tracks in upstream TPCs, and the TPC PID is
used on the highest-momentum negative track to isolate muons. This
selection has a 91\% purity for selecting $\nu_\mu$ charge-current
interactions.

The resulting data/Monte Carlo ratio, using the 2.88$\times10^{19}$
POT from the first T2K run, is
\[R = 1.036 \pm 0.028\mathrm{(stat)} ^{+0.044}_{-0.037}\mathrm{(det\ syst)} 
        \pm 0.038\mathrm{(phys\ model\ syst)}\]
This normalization reduces the flux uncertainty in the $\nu_e$
appearance measurement by 50\%.

\begin{figure}
  \centering
  \includegraphics[width=2.75in]{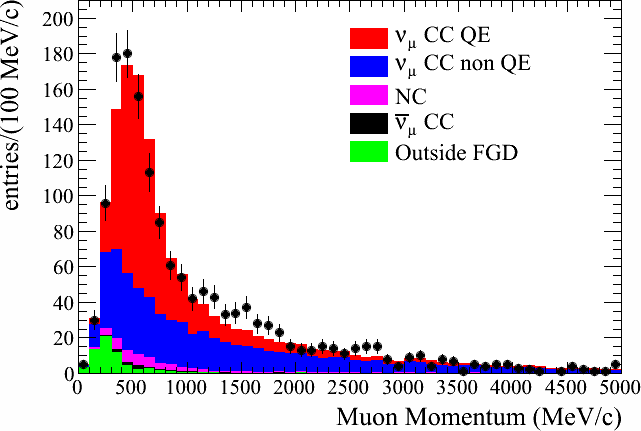}\includegraphics[width=2.75in]{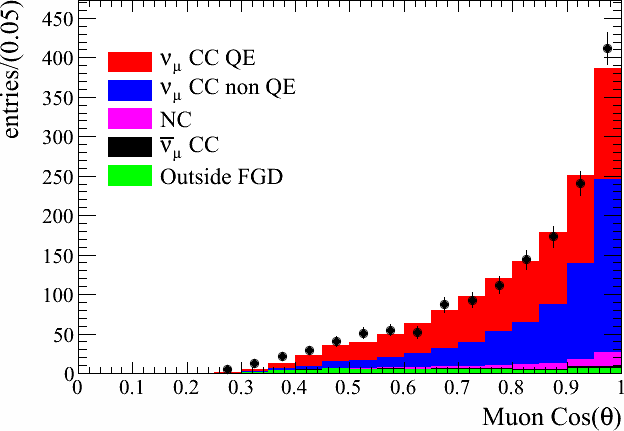}
  \caption{Distribution of muon momentum and angle in data from
    January-June 2010 and in Monte Carlo}
  \label{fig:mumom}
\end{figure}

Figure \ref{fig:mumom} shows the measured momentum and angular
distributions of the selected muons. In each plot, the points are the
data with statistical error bars, and the coloured histogram is the
Monte Carlo broken down by interaction type.

\subsection{Electron Neutrino Fraction Measurement}

The second measurement is of the electron neutrino fraction present in
the beam at its origin. This is a more difficult measurement;
charge-current muon neutrino interactions are a large fraction of the
beam events in ND280, while electron neutrinos have the large muon
contribution as background. 

While the selection for the electron neutrino candidates is quite
similar to that for the muon neutrinos, the extraction of the $\nu_e$
fraction must account for the large backgrounds. A likelihood fit with
empirical PDFs as a function of momentum for the signal and three
categories of background was used to extract the number of electron
neutrino interactions. 

\begin{figure}
  \centering
  \includegraphics[width=2.75in]{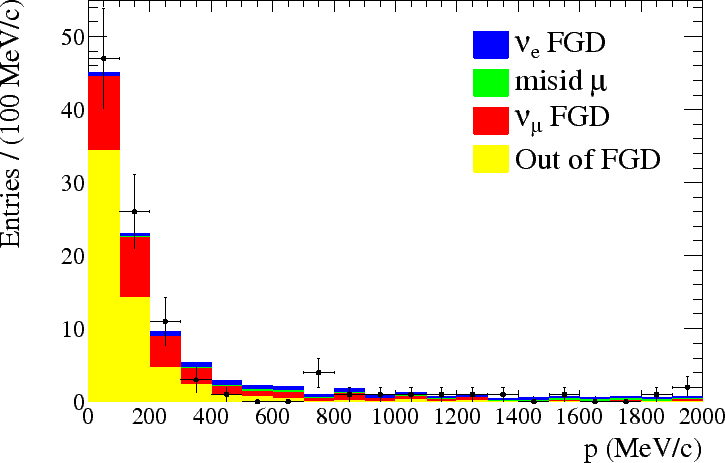}
  \caption{Momentum distribution of electron neutrino candidate events}
  \label{fig:efit}
\end{figure}

Figure \ref{fig:efit} shows the result of this fit. The resulting
$\nu_e$/$\nu_\mu$ ratio is $(1.0 \pm 0.7 (\mathrm{stat}) \pm 0.3
(\mathrm{syst}))$\%. To check the validity of the beam simulation, a
double ratio can be formed between data and Monte Carlo:
\[\frac{(\nu_e/\nu_\mu)_{data}}{(\nu_e/\nu_\mu)_{MC}} = 0.6 \pm
0.4 (\mathrm{stat}) \pm 0.2 (\mathrm{syst})\]
While this measurement is dominated by statistical uncertainties, it
does show a rough correspondence between the data and Monte Carlo. The
uncertainties are too large to constrain the background at Super-K,
but this provides a useful cross-check of our Monte Carlo simulation.

\section{Future Prospects}

Work is ongoing to extend the ND280 $\nu_\mu$ analysis to include a
selection of muon neutrino CCQE interactions. This, in turn, permits a
measurement of the neutrino spectrum, rather than just an overall
normalization. In addition, the additional data from Run 2 is being
incorporated into the analyses, pending consideration of the
differences in detector systematics.

The $\pi^0$ analysis is proceeding well. Reconstruction of decay
photons in the P0D is now possible to the precision necessary for
selection of $\pi^0$ decays. Incorporation of the Run 2 data and its
full complement of ECal modules will improve the $\pi^0$ analysis.

\section{Conclusion}

The T2K ND280 near detector is operational and is making significant
contributions to the oscillation measurements. Completed analyses
normalize the $\nu_\mu$ flux and measure the $\nu_e$ fraction of the
beam at its source. Improvements to these measurements are ongoing,
and work is underway to add a NC$\pi^0$ analysis to constrain that
important background.

\end{document}